\pdfoutput=1
\documentclass[prb,twocolumn,showpacs,aps]{revtex4}
\usepackage{graphicx,graphics,color,epsfig}
\usepackage{bm}
\usepackage{amsmath}
\usepackage{amssymb}
\usepackage{epstopdf}

\makeatletter

\newcommand{\Rmnum}[1]{\expandafter\@slowromancap\romannumeral #1@}
\makeatother

\usepackage{color}

\begin{document}

\title{Majorana zero modes in the hopping-modulated one-dimensional $p$-wave superconducting model}

\author{Yi Gao,$^{1}$ Tao Zhou,$^{2}$ Huaixiang Huang,$^{3}$ and Ran Huang,$^{1}$}
\affiliation{$^{1}$Department of Physics and Institute of Theoretical Physics,
Nanjing Normal University, Nanjing, 210023, China\\
$^{2}$College of Science, Nanjing University of Aeronautics and
Astronautics, Nanjing, 210016, China\\
$^{3}$Department of Physics, Shanghai University, Shanghai, 200444, China}

\begin{abstract}
We investigate the one-dimensional $p$-wave superconducting model with periodically modulated hopping and show that under time-reversal symmetry, the number of the Majorana zero modes (MZMs) strongly depends on the modulation period. If the modulation period is odd, there can be at most one MZM. However if the period is even, the number of the MZMs can be zero, one and two. In addition, the MZMs will disappear as the chemical potential varies. We derive the condition for the existence of the MZMs and show that the topological properties in this model are dramatically different from the one with periodically modulated potential.
\end{abstract}


\maketitle

\emph{Introduction}.---Recently, searching for Majorana fermions (MFs) in condensed matter systems has attracted much attention,~\cite{review1,review2,review3,review4} due to the following reasons. Firstly, MFs are their own antiparticles which were hypothesized to exist, but have long been elusive from detection. Secondly, because of their nonlocality and non-Abelian statistics, the zero-energy MFs, also called Majorana zero modes (MZMs), are proposed to be possible to realize fault tolerant topological quantum computation.~\cite{quntumcomputation1,quntumcomputation2,quntumcomputation3} Thus from both the fundamental and practical point of view, finding MFs is of great importance. In condensed matter systems, MFs can appear as quasiparticle excitations in topological superconductors and MZMs refer to the zero-energy in-gap excitations. There are several suggestions of physical systems that may support the MZMs,~\cite{kane,sato,sau,oreg,lutchyn,alicea} among which the one-dimensional $p$-wave superconducting (SC) model (also called the Kitaev model),~\cite{quntumcomputation1} due to its simplicity and elegance, is the most studied one. Possible realization of the Kitaev model includes quantum wires with a strong spin-orbit coupling (or topologically insulating wires subject to a Zeeman magnetic field) and in proximity to a superconductor.~\cite{lutchyn,oreg} In addition, it can also be realized in cold-atom systems.~\cite{sato,jiang}

Up to now, most of the theoretical works focus on ideal homogeneous \cite{quntumcomputation1} or potential-modulated Kitaev chains,~\cite{chenshu,degottardi,degottardi2} or Kitaev chains with longer-range hopping and pairing,~\cite{degottardi2,niu} or even quasi-one-dimensional Kitaev chains with a finite width.~\cite{potter,wakatsuki} Particularly in the periodically potential-modulated case,~\cite{chenshu,degottardi,degottardi2}, it was found that under time-reversal symmetry, the number of the MZMs can be at most one and if the potential vanishes at certain sites, then the MZM will be very robust and stable for arbitrary strength of the modulation. However, a very important problem unaddressed is the stability and fate of the MZMs under hopping modulation. In this work, we investigate the hopping-modulated one-dimensional $p$-wave SC model which is an extension of the original Kitaev model. We found that, under time-reversal symmetry, the number of the MZMs strongly depends on the period of the modulation. If the period is odd, there can be at most one MZM. However if the period is even, in some parameter regimes the number of the MZMs can be two. Furthermore, the MZMs will disappear as the chemical potential varies no matter the period is odd or even. Therefore the topological properties of the hopping-modulated model are drastically different from those of the potential-modulated one.

\emph{Method}.---We consider a one-dimensional Kitaev $p$-wave SC model where the hopping is periodically modulated, the Hamiltonian can be written as
\begin{eqnarray}
\label{h1}
H&=&\sum_{i}(-t_{i}c_{i}^{\dag}c_{i+1}+\Delta c_{i}c_{i+1}+H.c.)+Vc_{i}^{\dag}c_{i}\nonumber\\
&=&\frac{1}{2}C^{\dag}H_{BdG}C
=\frac{1}{2}C^{\dag}QQ^{\dag}H_{BdG}QQ^{\dag}C
=\frac{1}{2}\Phi^{\dag}\Lambda\Phi\nonumber\\
&=&\sum_{n=1}^{L}E_{n}\eta_{n}^{\dag}\eta_{n},
\end{eqnarray}
where $t_{i}=\cos(2\pi i\alpha+\delta)$ is the periodically modulated hopping integral. $\Delta\neq0$ is the $p$-wave SC pairing gap and $V$ is the chemical potential. Here $\alpha=p/q$ is a rational number with $p$ and $q$ being coprime integers. $C^{\dag}=(c_{1}^{\dag},\ldots,c_{L}^{\dag},c_{1},\ldots,c_{L})$ and $\Phi^{\dag}=(\eta_{1},\ldots,\eta_{L},\eta_{1}^{\dag},\ldots,\eta_{L}^{\dag})$, with $L$ being the number of the lattice sites. In addition,
\begin{eqnarray}
\label{E}
\Lambda&=&\begin{pmatrix}
-E_{L}& & & & & \\
 &\ddots& & & & \\
 & &-E_{1}& & & \\
 & & &E_{1}& & \\
 & & & &\ddots& \\
 & & & & &E_{L}
\end{pmatrix},
\end{eqnarray}
and $Q$ is a unitary matrix that diagonalizes $H_{BdG}$.

Defining Majorana operators $\gamma_{i}^{A}$ and $\gamma_{i}^{B}$ as
\begin{eqnarray}
\gamma_{i}^{A}&=&c_{i}^{\dag}+c_{i},\gamma_{i}^{B}=i(c_{i}-c_{i}^{\dag}),
\end{eqnarray}
then the quasiparticle operator $\eta_{n}^{\dag}$ can be expressed as
\begin{eqnarray}
\label{eta}
\eta_{n}^{\dag}&=&\frac{1}{2}\sum_{i=1}^{L}[\phi_{n,i}\gamma_{i}^{A}+\psi_{n,i}\gamma_{i}^{B}],
\end{eqnarray}
with $\phi_{n,i}$ and $\psi_{n,i}$ being the amplitudes of the MFs $\gamma_{i}^{A}$ and $\gamma_{i}^{B}$ in the $n$th eigenstate, respectively. If there exist MZMs, then none of the $E_{n}$ in Eq. (\ref{h1}) is zero under periodic boundary condition (PBC) while some of them become zero under open boundary condition (OBC) and the number of the MZMs is the number of the zero $E_{n}$.

Since $t_{i}$ is modulated with a period $q$ (the unit cell is enlarged by $q$ times), therefore under PBC, we have
\begin{eqnarray}
\label{h2}
H&=&\sum_{l=1}^{L/q}\sum_{s=1}^{q-1}(-t_{s}c_{s,l}^{\dag}c_{s+1,l}+\Delta c_{s,l}c_{s+1,l})\nonumber\\
& &-t_{q}c_{q,l}^{\dag}c_{1,l+1}+\Delta c_{q,l}c_{1,l+1}+H.c.\nonumber\\
& &+\sum_{l=1}^{L/q}\sum_{s=1}^{q}Vc_{s,l}^{\dag}c_{s,l}.
\end{eqnarray}
Using Fourier transform, $c_{s,l}^{\dag}=\sqrt{q/L}\sum_{k}c_{s,k}^{\dag}e^{ikql}$, $k\in(-\pi/q,\pi/q]$. In momentum space, we get
\begin{eqnarray}
\label{hk1}
H&=&\frac{1}{2}\sum_{k}\Psi_{k}^{\dag}H_{k}\Psi_{k},\nonumber\\
\Psi_{k}^{\dag}&=&(c_{1,k}^{\dag},\ldots,c_{q,k}^{\dag},c_{1,-k},\ldots,c_{q,-k}),\nonumber\\
H_{k}&=&\begin{pmatrix}
M_{k}&\Delta_{k}\\
\Delta_{k}^{\dag}&-M_{-k}^{T}
\end{pmatrix},
\end{eqnarray}
with the nonzero matrix elements of $M_{k}$ and $\Delta_{k}$ being $M_{k}^{s,s+1}=M_{k}^{s+1,s}=-t_{s}$ and
$\Delta_{k}^{s,s+1}=-\Delta_{k}^{s+1,s}=-\Delta$, for $s=1,\ldots,q-1$. $M_{k}^{s,s}=V$ for $s=1,\ldots,q$, while
$M_{k}^{q,1}=M_{k}^{1,q*}=-t_{q}e^{-ikq}$ and
$\Delta_{k}^{q,1}=-\Delta_{k}^{1,q*}=-\Delta e^{-ikq}$.

Since we assume $t_{i}$ and $\Delta$ in Eq. (\ref{h1}) is real (up to a global phase) throughout the paper, thus $H_{k}$ respects the time-reversal, particle-hole and chiral symmetries and it can be unitarily transformed to an off-diagonal matrix as \cite{wakatsuki,Z,xiong}
\begin{eqnarray}
\label{u}
UH_{k}U^{\dag}&=&\begin{pmatrix}
0&A_{k}\\
A_{-k}^{T}&0
\end{pmatrix},A_{k}=M_{k}+\Delta_{k},\nonumber\\
U&=&e^{-i\pi\tau_{y}/4}=\frac{1}{\sqrt{2}}\begin{pmatrix}
I&-I\\
I&I
\end{pmatrix},
\end{eqnarray}
here $\tau_{y}$ is a Pauli matrix acting on the particle-hole space. Then the system belongs to the class BDI which is characterized by the $\mathbb{Z}$ index while the number of the MZMs can be represented by $W$ which is calculated through
\begin{eqnarray}
\label{w}
W&=&\frac{-i}{\pi}\int_{k=0}^{k=\pi/q}\frac{dz_{k}}{z_{k}},\nonumber\\
z_{k}&=&e^{i\theta_{k}}=\frac{Det(A_{k})}{|Det(A_{k})|}.
\end{eqnarray}
In fact, $W$ just counts how many times the determinant of $A_{k}$ crosses the imaginary axis as $k$ evolves from $0$ to $\pi/q$.

\begin{figure}
\includegraphics[width=1\linewidth]{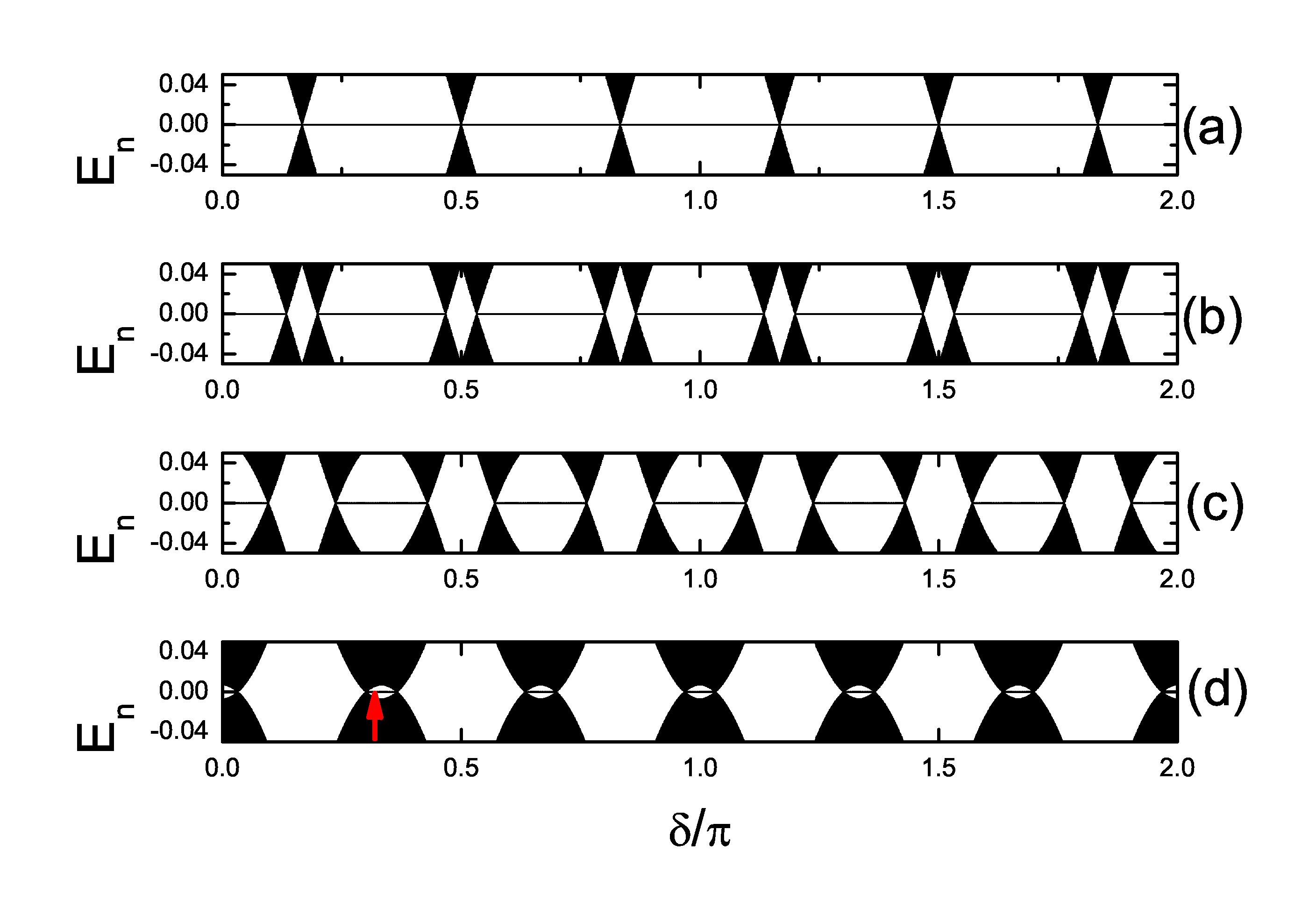}
 \caption{\label{1/3_En} The energy spectra for $\alpha=1/3$ under OBC. Here $\Delta=1$ and $L=1632$. (a), (b), (c) and (d) are for $V=0$, $0.1$, $0.2$ and $0.3$, respectively. Only $|E_{n}|<0.05$ are plotted.}
\end{figure}

\begin{figure}
\includegraphics[width=1\linewidth]{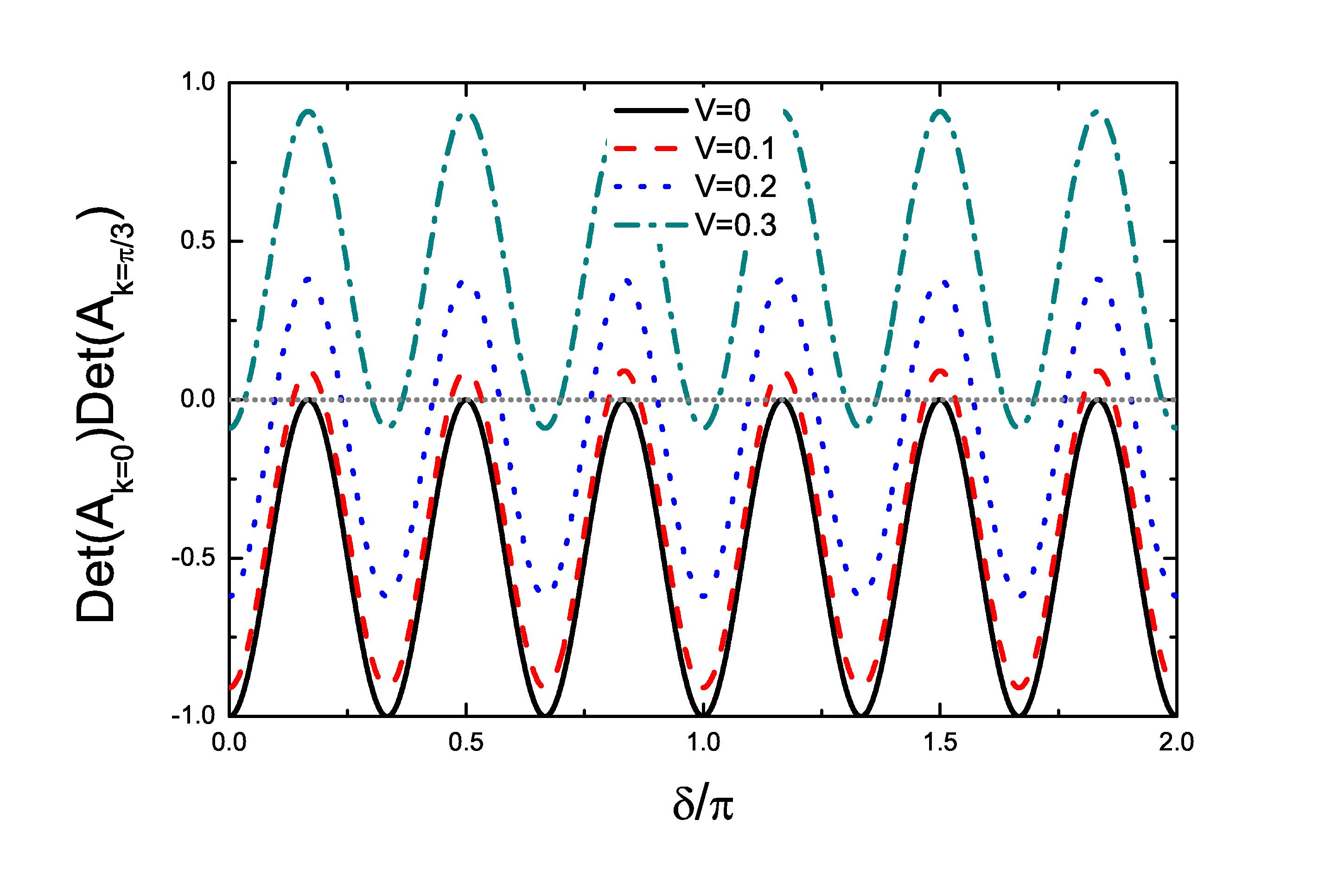}
 \caption{\label{1/3_W} (Color online) $Det(A_{k=0})Det(A_{k=\pi/3})$ as a function of $\delta$ and $V$, for $\alpha=1/3$ under PBC. The black solid, red dashed, blue dotted and green dash dotted lines are for $V=0$, $0.1$, $0.2$ and $0.3$, respectively. The gray dotted line denotes $Det(A_{k=0})Det(A_{k=\pi/3})=0$. Here $\Delta=1$ and $L=1632$.}
\end{figure}

\begin{figure}
\includegraphics[width=1\linewidth]{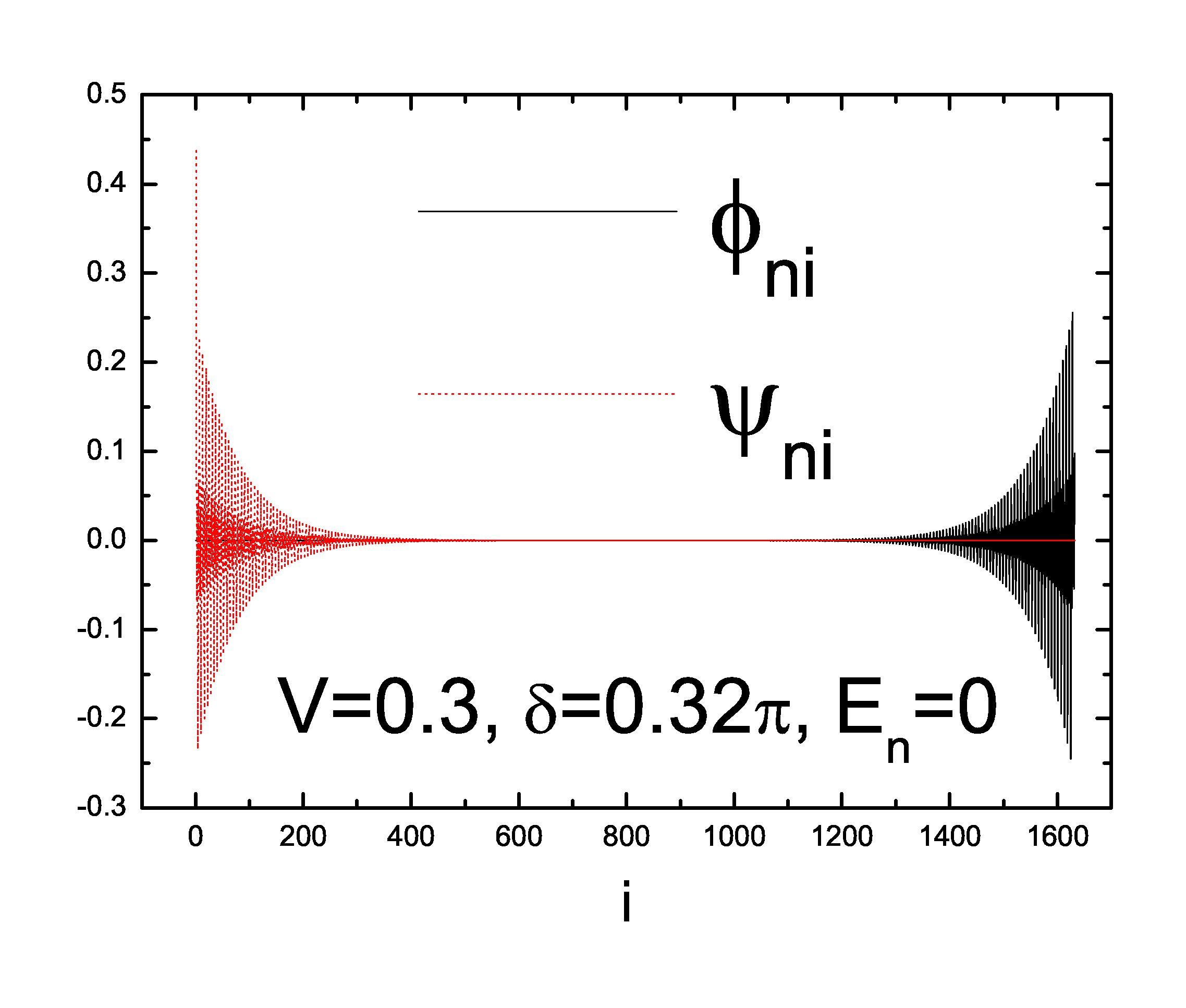}
 \caption{\label{1/3_distribution} (Color online) The distribution of the zero-mode MFs along the one-dimensional lattice for $\alpha=1/3$ under OBC. Here $\Delta=1$, $L=1632$, $V=0.3$ and $\delta=0.32\pi$, as denoted by the red arrow in Fig. \ref{1/3_En}(d).}
\end{figure}

\begin{figure}
\includegraphics[width=1\linewidth]{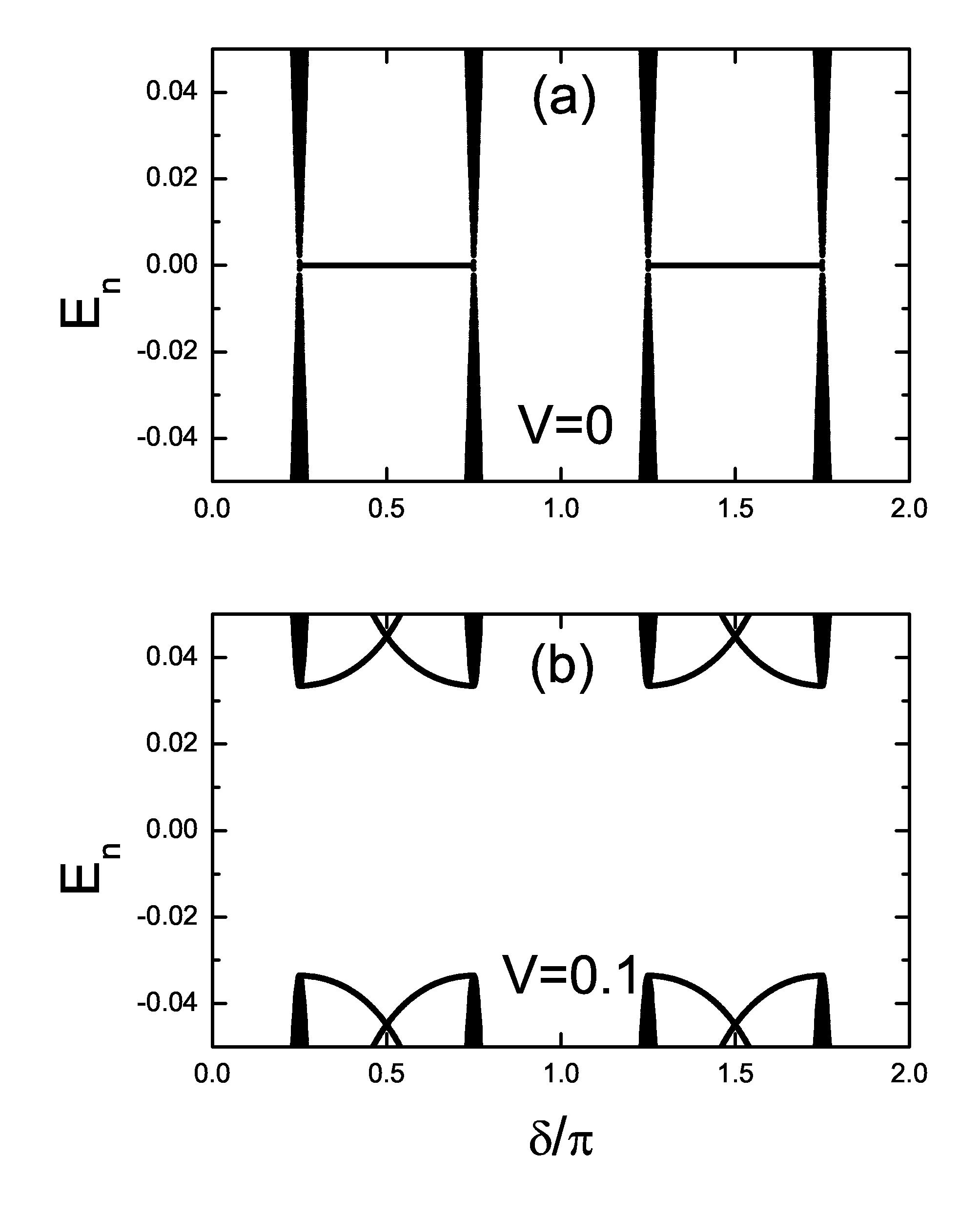}
 \caption{\label{1/4_En} The energy spectra for $\alpha=1/4$ under OBC. Here $\Delta=1$ and $L=1632$. (a) $V=0$. (b) $V=0.1$. Only $|E_{n}|<0.05$ are plotted.}
\end{figure}

\begin{figure}
\includegraphics[width=1\linewidth]{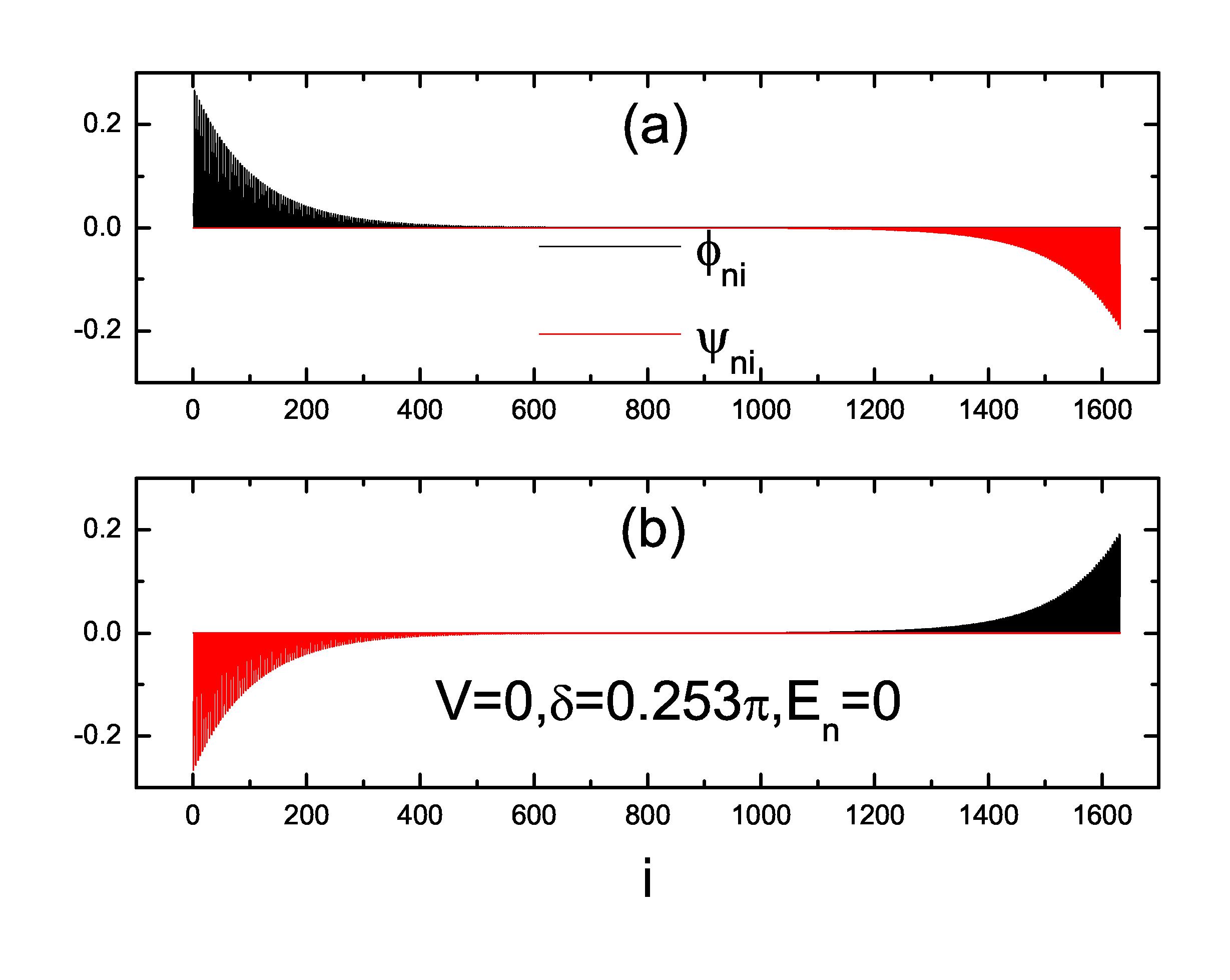}
 \caption{\label{1/4_distribution} (Color online) The distribution of the zero-mode MFs along the one-dimensional lattice for $\alpha=1/4$ under OBC. Here $\Delta=1$, $L=1632$, $V=0$ and $\delta=0.253\pi$. (a) The eigenstate corresponding to $E_{1}=0$. (b) The eigenstate corresponding to $E_{2}=0$. }
\end{figure}

\emph{Results}.---First we consider the $\alpha=1/2$ case, where $t_{1}=-t_{2}=-\cos\delta$. Under PBC, we have
\begin{eqnarray}
M_{k}&=&\begin{pmatrix}
V&-(t_{1}+t_{2}e^{2ik})\\
-(t_{1}+t_{2}e^{-2ik})&V
\end{pmatrix},\nonumber\\
\Delta_{k}&=&\begin{pmatrix}
0&-\Delta(1-e^{2ik})\\
\Delta(1-e^{-2ik})&0
\end{pmatrix},\nonumber\\
Det(A_{k})&=&V^{2}-t_{1}^{2}-t_{2}^{2}+2\Delta^{2}-2(t_{1}t_{2}+\Delta^{2})\cos2k\nonumber\\
& &+2i\Delta(t_{1}+t_{2})\sin2k\nonumber\\
&=&V^{2}+4(\Delta^{2}-\cos^{2}\delta)\sin^{2}k.
\end{eqnarray}
Since $Det(A_{k})$ is real, $W$ must be zero otherwise $Det(A_{k})$ will be zero for some $k$ (which means the bulk energy gap vanishes), therefore, there can be no MZMs.

Generally, the periodic modulation can take many forms. For arbitrary $t_{1}$ and $t_{2}$ ($t_{1}\neq-t_{2}$), we have
\begin{eqnarray}
Det(A_{k=0})&=&V^{2}-(t_{1}+t_{2})^{2},\nonumber\\
Det(A_{k=\pi/2})&=&V^{2}-(t_{1}-t_{2})^{2}+4\Delta^{2}.
\end{eqnarray}
If $V\neq0$ and $Det(A_{k=0})Det(A_{k=\pi/2})<0$, then $Det(A_{k})$ will cross the imaginary axis once as $k$ changes from $0$ to $\pi/2$. In this case, $W=1$ and one MZM
exists. Interestingly, at $V=0$, we have
\begin{eqnarray}
A_{k}&=&\begin{pmatrix}
0&A_{1k}\\
A_{2k}&0
\end{pmatrix},\nonumber\\
A_{1k}&=&-[t_{1}+t_{2}e^{2ik}+\Delta(1-e^{2ik})],\nonumber\\
A_{2k}&=&-[t_{1}+t_{2}e^{-2ik}-\Delta(1-e^{-2ik})].
\end{eqnarray}
In this case, the system can be divided into two separated subsystems $A_{1k}$ and $A_{2k}$. As $k$ evolves from $0$ to $\pi/2$, we have
\begin{eqnarray}
A_{1k=0}A_{1k=\pi/2}&=&(\Delta+t_{1})^{2}-(\Delta-t_{2})^{2},\nonumber\\
A_{2k=0}A_{2k=\pi/2}&=&(\Delta-t_{1})^{2}-(\Delta+t_{2})^{2}.
\end{eqnarray}
Therefore, if both $A_{1k=0}A_{1k=\pi/2}$ and $A_{2k=0}A_{2k=\pi/2}$ are less than zero, then two MZMs will show up in this case and the existence of these two MZMs has been numerically verified (for example, $t_{1}=0.5$, $t_{2}=-0.8$ and $\Delta=0.5$).

For $\alpha=1/3$, we have
\begin{eqnarray}
M_{k}&=&\begin{pmatrix}
V&-t_{1}&-t_{3}e^{3ik}\\
-t_{1}&V&-t_{2}\\
-t_{3}e^{-3ik}&-t_{2}&V
\end{pmatrix},\nonumber\\
\Delta_{k}&=&\begin{pmatrix}
0&-\Delta&\Delta e^{3ik}\\
\Delta&0&-\Delta\\
-\Delta e^{-3ik}&\Delta&0
\end{pmatrix},
\end{eqnarray}
where $t_{1}=\cos(2\pi/3+\delta)$, $t_{2}=\cos(4\pi/3+\delta)$ and $t_{3}=\cos\delta$. In this case,
\begin{eqnarray}
Det(A_{k})&=&[V(-3+6\Delta^{2}+2V^{2})-\cos3k\cos3\delta\nonumber\\
& &+i\Delta(4\Delta^{2}-3)\sin3k]/2,
\end{eqnarray}
and $Det(A_{k=0,\pi/3})=[\mp\cos3\delta+V(-3+6\Delta^{2}+2V^{2})]/2$ ($-$ and $+$ are for $k=0$ and $\pi/3$, respectively). If $Det(A_{k=0})Det(A_{k=\pi/3})<0$, then $Det(A_{k})$ will cross the imaginary axis exactly once as $k$ evolves from $0$ to $\pi/3$, which means that $W=1$ and one MZM exists in this case [$E_{1}$ in Eq. (\ref{E}) is zero under OBC]. On the contrary, $Det(A_{k=0})Det(A_{k=\pi/3})=0$ means that the bulk energy gap vanishes and $Det(A_{k=0})Det(A_{k=\pi/3})>0$ means that $Det(A_{k})$ will not cross the imaginary axis as $k$ evolves from $0$ to $\pi/3$, in both cases $W=0$ and no MZMs exist. Specifically, for $\delta=(2m+1)\pi/6$ with $m$ being an integer, we have $\cos3\delta=0$, therefore $Det(A_{k=0})Det(A_{k=\pi/3})\geqslant0$ and no MZMs can exist, irrespective of the values of $\Delta$ and $V$. For example, we set $\Delta=1$ and $L=1632$. In Figs. \ref{1/3_En} and \ref{1/3_W} we plot the energy spectra under OBC and $Det(A_{k=0})Det(A_{k=\pi/3})$ under PBC, respectively. As we can see, at $V=0$, MZM exists for any $\delta$, except for $\delta=\pi/6,\pi/2,5\pi/6,7\pi/6,3\pi/2,11\pi/6$ where the bulk energy gap closes. As $V$ increases, for some $\delta$, MZM vanishes and as $\delta$ evolves from $0$ to $2\pi$, topologically trivial (without MZM) and nontrivial (with one MZM) phases appear in turn. A typical distribution of the zero-mode MFs is shown in Fig. \ref{1/3_distribution} and we can see that the two MFs $\gamma_{i}^{A}$ and $\gamma_{i}^{B}$ are well separated in real space and are located at the left and right ends, respectively while the actual decay length of these two MFs increases/decreases as the bulk energy gap decreases/increases. Finally when $V>0.32$ where the condition $|V(-3+6\Delta^{2}+2V^{2})|>1$ is satisfied, MZM disappears for any $\delta$ and there is only topologically trivial phase and indeed for $\delta=(2m+1)\pi/6$ with $m$ being an integer, MZMs do not exist for any $V$.

For $\alpha=1/4$, we have $t_{1}=-t_{3}=-\sin\delta$ and $t_{2}=-t_{4}=-\cos\delta$. In this case,
\begin{eqnarray}
M_{k}&=&\begin{pmatrix}
V&-t_{1}&0&-t_{4}e^{4ik}\\
-t_{1}&V&-t_{2}&0\\
0&-t_{2}&V&-t_{3}\\
-t_{4}e^{-4ik}&0&-t_{3}&V
\end{pmatrix},\nonumber\\
\Delta_{k}&=&\begin{pmatrix}
0&-\Delta&0&\Delta e^{4ik}\\
\Delta&0&-\Delta&0\\
0&\Delta&0&-\Delta\\
-\Delta e^{-4ik}&0&\Delta&0
\end{pmatrix},\nonumber\\
Det(A_{k})&=&[3-8\Delta^{2}+8\Delta^{4}+8(2\Delta^{2}-1)V^{2}+4V^{4}\nonumber\\
& &+\cos4\delta+\cos4k(-1+8\Delta^{2}-8\Delta^{4}\nonumber\\
& &+\cos4\delta)]/4.
\end{eqnarray}
At first glance, since $Det(A_{k})$ is real, thus, similar to the $\alpha=1/2$ case, there should be no MZMs. However, this is not true at $V=0$. In the following we set $\Delta=1$ and $L=1632$ as an example. As we can see from Fig. \ref{1/4_En}, indeed at $V=0.1$, there are no MZMs. However at $V=0$, MZMs exist for $\pi/4<\delta<3\pi/4$ and $5\pi/4<\delta<7\pi/4$. These MZMs are doubly degenerate [both $E_{1}$ and $E_{2}$ in Eq. (\ref{E}) are zero under OBC] and the distribution of the zero-mode MFs is shown in Fig. \ref{1/4_distribution}. The existence of these two MZMs can be explained as follows. At $V=0$, we found that the eigenvalues of $H_{k}$ in Eq. (\ref{hk1}) are doubly degenerate, therefore $H_{k}$ can be divided into two independent subsystems by a unitary transformation as
\begin{eqnarray}
\label{subsystems}
PH_{k}P^{\dag}&=&\begin{pmatrix}
H_{1k}&0\\
0&H_{2k}
\end{pmatrix},
\end{eqnarray}
here both $H_{1k}$ and $H_{2k}$ are $4\times4$ matrices while their eigenvalues are exactly the same. The unitary matrix $P$ can be written as
\begin{eqnarray}
P&=&\frac{1}{2}\begin{pmatrix}
1&1&0&0&1&-1&0&0\\
0&0&1&1&0&0&1&-1\\
1&-1&0&0&1&1&0&0\\
0&0&1&-1&0&0&1&1\\
1&1&0&0&-1&1&0&0\\
0&0&1&1&0&0&-1&1\\
-1&1&0&0&1&1&0&0\\
0&0&-1&1&0&0&1&1
\end{pmatrix},
\end{eqnarray}
and
\begin{eqnarray}
H_{jk}&=&\begin{pmatrix}
M_{jk}&\Delta_{jk}\\
-\Delta_{jk}&-M_{jk}
\end{pmatrix},\text{$j=1,2$},\nonumber\\
Det(A_{1k})&=&[Det(A_{2k})]^{*}\nonumber\\
&=&(\Delta^{2}-\sin^{2}\delta)-(\Delta^{2}-\cos^{2}\delta)e^{4ik}.
\end{eqnarray}
At $k=0$, $Det(A_{1k})=Det(A_{2k})=\cos2\delta$ while at $k=\pi/4$, $Det(A_{1k})=Det(A_{2k})=2\Delta^2-1$. If $\cos2\delta(2\Delta^2-1)<0$, then both $Det(A_{1k})$ and $Det(A_{2k})$ will cross the imaginary axis exactly once as $k$ evolves from $0$ to $\pi/4$, indicating that there exists one MZM in each subsystem and the number of the MZMs for the whole system is two. Therefore for $\alpha=1/4$, at $V=0$, the number of the MZMs are either two or zero while at $V\neq0$, there are no MZMs. For general $t_{i}$ ($i=1,\ldots,4$), at $V\neq0$, $Det(A_{k})$ may not be real and there may exist one MZM. However at $V=0$, the system can still be divided into two subsystems. In this case, if the conditions $[(\Delta+t_{1})(\Delta+t_{3})]^{2}-[(\Delta-t_{2})(\Delta-t_{4})]^{2}<0$ and $[(\Delta-t_{1})(\Delta-t_{3})]^{2}-[(\Delta+t_{2})(\Delta+t_{4})]^{2}<0$ are satisfied simultaneously, there will be two MZMs.

Furthermore we found that, for general periodic modulation, if the period $q$ is odd, then the number of the MZMs is either zero or one. On the other hand, if $q$ is even, then at $V\neq0$, the number of the MZMs is still zero or one. However at $V=0$, the system can always be divided into two independent subsystems and if the conditions
\begin{eqnarray}
\label{condition1}
[(\Delta+t_{1})(\Delta+t_{3})\ldots(\Delta+t_{q-1})]^{2}\nonumber\\
-[(\Delta-t_{2})(\Delta-t_{4})\ldots(\Delta-t_{q})]^{2}<0,
\end{eqnarray}
and
\begin{eqnarray}
\label{condition2}
[(\Delta-t_{1})(\Delta-t_{3})\ldots(\Delta-t_{q-1})]^{2}\nonumber\\
-[(\Delta+t_{2})(\Delta+t_{4})\ldots(\Delta+t_{q})]^{2}<0,
\end{eqnarray}
are simultaneously satisfied, there will be two MZMs. However in this case, the subsystems cannot be further divided, making the maximal number of the MZMs be two in the whole system.

\emph{Summary}.---In summary, we have studied the number of the MZMs and their stability in the hopping-modulated one-dimensional $p$-wave SC model. We found that the former strongly depends on the period of the modulation. If the period $q$ is odd, there can be at most one MZM in the system while for an even $q$, the number of the MZMs can be zero, one and two. The existence of two MZMs can occur only at $V=0$, since in this case, $A_{k}$ in Eq. (\ref{u}) can always be divided into two independent sub-matrices by a unitary transformation as
\begin{eqnarray}
SA_{k}S^{T}&=&\begin{pmatrix}
0&A_{1k}\\
A_{2k}&0
\end{pmatrix},\nonumber\\
S_{(j+1)/2,j}&=&1,\text{for $j=1,3,5,\ldots,q-1$},\nonumber\\
S_{(j+q)/2,j}&=&1,\text{for $j=2,4,6,\ldots,q$}.
\end{eqnarray}
At certain conditions, there exists one MZM in each subsystem and the number of the MZMs for the whole system is two. For the specific modulation form we considered [$t_{i}=\cos(2\pi i\alpha+\delta)$ with $\alpha=p/q$], only at $q=4n$ ($n=1,2,3,\ldots$) can Eqs. (\ref{condition1}) and (\ref{condition2}) be simultaneously satisfied, therefore only in this case can there exist two MZMs. Furthermore, the MZMs will vanish as the chemical potential $V$ varies. In the periodically potential-modulated model considered in Refs. \onlinecite{chenshu}, \onlinecite{degottardi} and \onlinecite{degottardi2}, when the time-reversal symmetry is present, there can be at most one MZM and if the potential vanishes at certain sites, then the MZM will be very robust and stable for arbitrary strength of the modulation. Clearly this is not the case in the periodically hopping-modulated model, therefore the topological properties differ drastically between these two models.

We thank Q. H. Wang, Y. Xiong and P. Q. Tong for helpful discussions. This work was supported by NSFC (Grants No. 11204138 and No. 11374005), NSF of Jiangsu Province of China (Grant No. BK2012450), NSF of Shanghai (Grant No.13ZR1415400), SRFDP (Grant No. 20123207120005) and NCET (Grant No. NCET-12-0626).

\end{document}